\documentclass[journal,twoside,web]{ieeecolor}
\usepackage{tmi}
\usepackage{cite}
\usepackage{amsfonts}
\usepackage{graphicx}
\usepackage{textcomp}
\def\BibTeX{{\rm B\kern-.05em{\sc i\kern-.025em b}\kern-.08em
    T\kern-.1667em\lower.7ex\hbox{E}\kern-.125emX}}
\usepackage{graphicx}
\usepackage{comment}
\usepackage{amsmath,amssymb} 
\usepackage{color}
\usepackage{amsmath}
\usepackage{amssymb}
\usepackage{latexsym}
\usepackage{url}
\usepackage{cite}
\usepackage{relsize}
\usepackage{multirow}
\usepackage{tabularx}
\usepackage{array}
\usepackage{afterpage}
\usepackage{ifthen}
\usepackage{algpseudocode,algorithmicx}
\usepackage{subfigure}
\usepackage[keeplastbox]{flushend}
\usepackage{epstopdf}
\usepackage{stackengine}
\usepackage{gensymb}
\usepackage{epsfig}
\usepackage{ mathrsfs }
\usepackage{tablefootnote}
\usepackage{ stmaryrd }

\usepackage{soul}
\makeatletter
\AtBeginDocument{\let\hl\@firstofone}
\makeatother

\usepackage[ruled,vlined]{algorithm2e}
\usepackage{pifont}

\newcolumntype{L}[1]{>{\raggedright\let\newline\\\arraybackslash\hspace{0pt}}m{#1}}
\newcolumntype{C}[1]{>{\centering\let\newline\\\arraybackslash\hspace{0pt}}m{#1}}
\newcolumntype{R}[1]{>{\raggedleft\let\newline\\\arraybackslash\hspace{0pt}}m{#1}}
\usepackage{supertabular}
\usepackage{ dsfont }

\DeclareUnicodeCharacter{2212}{-}

\usepackage[bookmarks=false, linkcolor=blue, urlcolor=blue, citecolor=blue]{hyperref} 
\usepackage{booktabs}
\usepackage{makecell}
\usepackage{hhline}
\usepackage{courier}
\usepackage{lipsum}

\usepackage{amsmath,amsfonts,bm} 

\usepackage{textcomp}
\def\BibTeX{{\rm B\kern-.05em{\sc i\kern-.025em b}\kern-.08em
    T\kern-.1667em\lower.7ex\hbox{E}\kern-.125emX}}
\begin{document}
\title{Light-weight Deformable Registration using \\ Adversarial Learning with Distilling Knowledge}
\author{Minh Q. Tran, Tuong Do, Huy Tran, Erman Tjiputra, Quang D. Tran, Anh Nguyen
\thanks{Minh Q. Tran, Tuong Do, Huy Tran, Erman Tjiputra, Quang D. Tran are with AIOZ, Singapore {\{minh.quang.tran, tuong.khanh-long.do, onekyc, erman.tjiputra, quang.tran\}@aioz.io}.}
\thanks{Anh Nguyen is with The Hamlyn Centre for Robotic Surgery, Imperial College London, UK (email: a.nguyen@imperial.ac.uk).}
\thanks{Two first authors contribute equally.}
\thanks{Corresponding author: Anh Nguyen.}
}

\maketitle
\begin{abstract}
Deformable registration is a crucial step in many medical procedures such as image-guided surgery and radiation therapy. Most recent learning-based methods focus on improving the accuracy by optimizing the non-linear spatial correspondence between the input images. Therefore, these methods are computationally expensive and require modern graphic cards for real-time deployment. In this paper, we introduce a new Light-weight Deformable Registration network that significantly reduces the computational cost while achieving competitive accuracy. In particular, we propose a new adversarial learning with distilling knowledge algorithm that successfully leverages meaningful information from the effective but expensive teacher network to the student network. We design the student network such as it is light-weight and well suitable for deployment on a typical CPU. The extensively experimental results on different public datasets show that our proposed method achieves state-of-the-art accuracy while significantly faster than recent methods. We further show that the use of our adversarial learning algorithm is essential for a time-efficiency deformable registration method. Finally, our source code and trained models are available at \url{https://github.com/aioz-ai/LDR_ALDK}.
\end{abstract}
\begin{IEEEkeywords}
Adversarial Learning, Deformable Registration, Knowledge Distillation, Light-weight Network, Time Efficiency.
\end{IEEEkeywords}

\section{Introduction}
Medical image registration is the process of systematically placing separate medical images in a common frame of reference so that the information they contain can be effectively integrated or compared~\cite{bashiri2019Manifold}.
Applications of image registration include combining images of the same subject from different modalities, aligning temporal sequences of images to compensate for the motion of the subject between scans, aligning images from multiple subjects in cohort studies, or navigating with image guidance during interventions~\cite{blendowski2020ShapeEncodeDecode, goubran2019MicroscopyMRI,kim2019CycleConsistent,de2019Affine,fan2019birnet,nguyen2020end,kundrat2021mr, drobny2018registration,do2021multiple,huang2021self,maintz1998survey}. Since many organs do deform substantially while being scanned, the rigid assumption can be violated as a result of scanner-induced geometrical distortions that differ between images. Therefore, performing deformable registration is an essential step in many medical procedures.

\begin{figure}
\centering
    \subfigure[]{\includegraphics[width=0.45\textwidth]{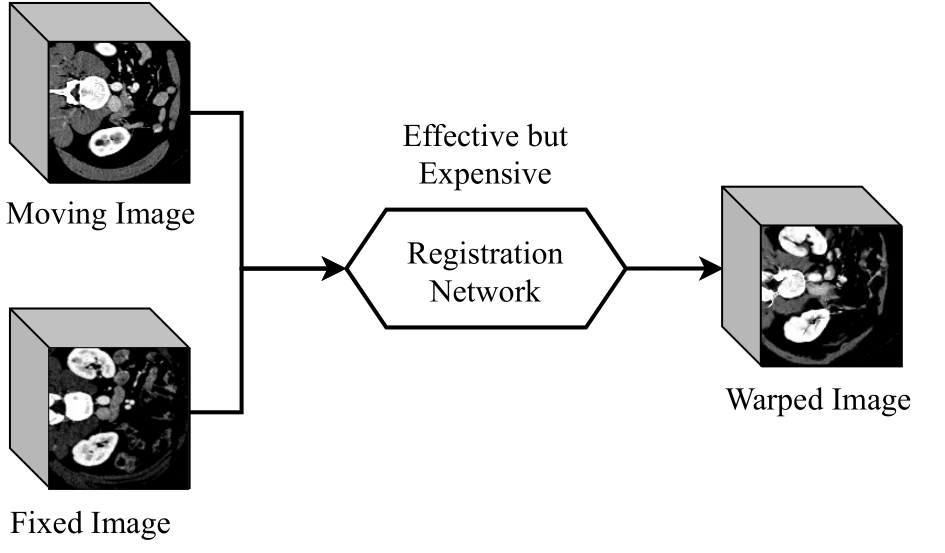}} 
    \subfigure[]{\includegraphics[width=0.45\textwidth, keepaspectratio=true]{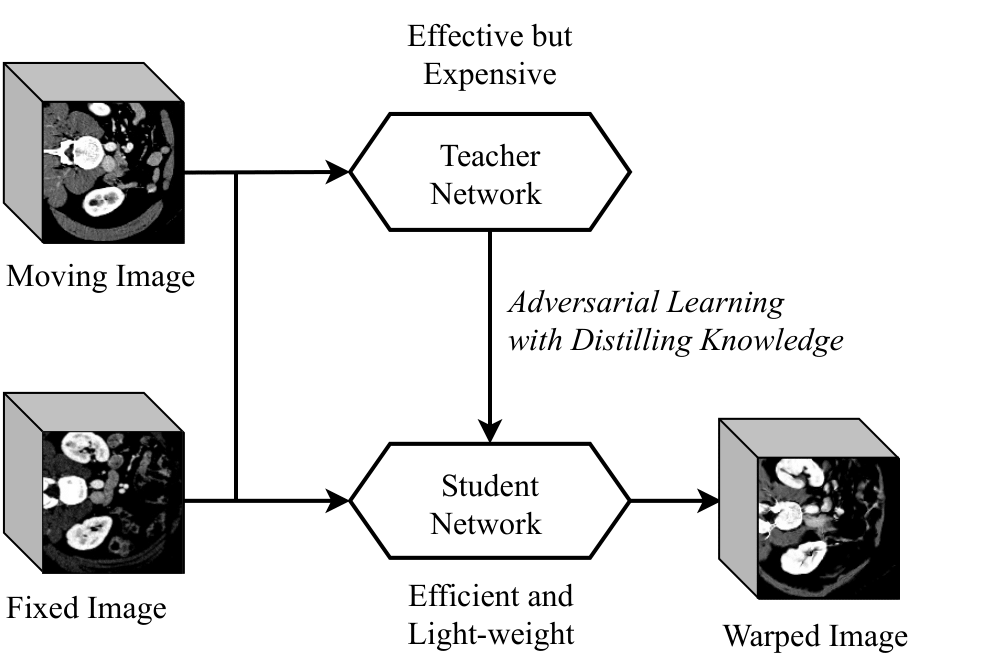}} 
    \caption{
    Comparison between typical deep learning-based methods for deformable registration (\textbf{a}) and our approach using adversarial learning with distilling knowledge for deformable registration (\textbf{b}). In our work, the expensive Teacher Network is used only in training; the Student Network is light-weight and inherits helpful knowledge from the Teacher Network via our Adversarial Learning algorithm. Therefore, the Student Network has high inference speed, while achieving competitive accuracy.}
    \vspace{-3ex}
    \label{fig:intro}
\end{figure}

Recently, learning-based methods have become popular to tackle the problem of deformable registration. These methods can be split into two groups: \textit{(i)} supervised methods that rely on the dense ground-truth flows obtained by either traditional algorithms or simulating intra-subject deformations \cite{SteeredCNN,yang2017quicksilver,krebs2017robustAction, sokooti2017nonrigid}. Although these works achieve state-of-the-art performance, they require a large amount of manually labeled training data, which are expensive to obtain; and \textit{(ii)} unsupervised learning methods that use a similarity measurement between the moving and the fixed image to utilize a large amount of unlabelled  data~\cite{VoxelMorph, RCN, VTN}. These unsupervised methods achieve competitive results in comparison with supervised methods. However, their deformations are reconstructed without the direct ground-truth guidance, hence leading to the limitation of leveraging learnable information~\cite{VTN}. Furthermore, recent unsupervised methods all share an issue of great complexity as the network parameters increase significantly when multiple progressive cascades are taken into account~\cite{RCN}. This leads to the fact that these works can not achieve real-time performance during inference while requiring intensively computational resources when deploying. 
 
In practice, there are many scenarios when medical image registration are needed to be fast $\--$
consider matching preoperative and intra-operative images during surgery, interactive change detection of CT or MRI data for a radiologist, deformation compensation or 3D alignment of large histological slices for a pathologist, or processing large amounts of images from high-throughput imaging methods \cite{kybic2018fast}.
Besides, in many image-guided robotic interventions, performing real-time deformable registration is an essential step to register the images and deal with organs that deform substantially~\cite{ge2019landmark}.
Economically, the development of a CPU-friendly solution for deformable registration will significantly reduce the instrument costs equipped for the operating theatre, as it does not require GPU or cloud-based computing servers, which are costly and consume much more power than CPU. This will benefit patients in low- and middle-income countries, where they face limitations in local equipment, personnel expertise, and
budget constraints infrastructure~\cite{mollura2020artificial}. Therefore, design an efficient model which is fast and accurate for deformable registration is a crucial task and worth for study in order to improve a variety of surgical interventions.

In this paper, we propose a new deformable registration method that can achieve competitive results with other state-of-the-art approaches, while significantly decreasing the network parameters and inference time (Fig. \ref{fig:intro}). Our principal contribution is a robust adversarial learning algorithm that leverages distilled knowledge from a teacher network to a light-weight student model. Our observation is based on the fact that although the teacher network achieves state-of-the-art performance, it has a large number of parameters and computationally expensive. Therefore, we only employ the teacher network during the training period then use our adversarial learning algorithm to leverage meaningful knowledge of the teacher network to the student network. Apart from the adversarial learning algorithm, we design a new light-weight student network which significantly reduces the inference time, making our deformable registration framework more suitable for real-world medical procedures. 

Our contributions can be summarised as follow:

\begin{itemize}
    \item We propose a new adversarial learning with distilling knowledge algorithm that can perform deformable registration effectively and timely. 
    \item We design a new light-weight network to act as a student network, which allows our system to achieve fast inference time using just a typical CPU.
    \item We extensively evaluate our method on several datasets to validate the results. Our source and trained models are publicly available for reproducibility.
\end{itemize}

Next, we review the related work in Section~\ref{sec_rw}. We then describe our adversarial learning with distilling knowledge algorithm for light-weight deformable registration in Section~\ref{learning_strat}. In Section~\ref{sec_exps}, we present extensively experimental results and compare our method with recent approaches. Finally, we discuss and conclude the paper in Section~\ref{sec_con}.

\section{Related Work}
\label{sec_rw}

\textbf{Deformable Registration}.
Medical image registration is a popular research topic in medical imaging~ \cite{bashiri2019Manifold, blendowski2020ShapeEncodeDecode, goubran2019MicroscopyMRI,kim2019CycleConsistent,de2019Affine,fan2019birnet,niethammer2019metric,SyN,B-spline}. Recently, lightweight methods for medical image registration have explored in many tasks including: fluid registration \cite{bro1996fast}, surface registration \cite{granger2002multi}, symmetric  registration \cite{reaungamornrat2016mind,ofverstedt2019fast}, electron microscopy registration \cite{zhou2019fast}, rigid registration \cite{schneider2012real}, elastic registration of soft tissues \cite{peterlik2018fast}, x-ray/echo registration \cite{hatt2016real} and thoracic 4D CT registration \cite{sentker2018gdl}. To embark upon, there are many approaches that try to find an optimal transformation. Myriad works develop tools such as FAIR \cite{FAIR}, ANTs \cite{ANTs}, and Elastix \cite{Elastix}, which iteratively update the parameters of the defined alignment objective. These optimization procedures are time consuming for practically clinical applications.
With the recent rise of deep learning, supervised methods are widely used in medical image registration~\cite{litjens2017survey,sokooti2017nonrigid,Flownet,yang2017quicksilver,krebs2017robustAction, hu2018WeaklyMultiImgRegis, hu2018label, ge2019landmark}. Despite their adequate performance, they demand copious ground-truth alignment or synthetic data that have to be generated with careful designs to resemble the real ones. 

To overcome the shortcoming of the supervised approaches, unsupervised methods are introduced for deformable registration~\cite{VoxelMorph,VTN,RCN,DLIR,shen2019regionMetricMapping, shen2019networks, kuang2019faim, dalca2018unsupervised}. Specifically, VoxelMorph\cite{VoxelMorph} predicts a dense deformation using deconvolutional layers \cite{deconvolutional}, whereas VTN \cite{VTN} proposes an end-to-end framework by substituting the traditional affine stage by the one utilizing in a convolutional neural network. 
By building on these two base networks, RCN \cite{RCN} outperforms state-of-the-art methods by presenting recursive cascaded networks, in which every cascade learns to perform a progressive deformation for the current warped image.
However, compared with the supervised methods, deformations are reconstructed without the direct ground-truth guidance, hence the learnable information of the network is limited~\cite{VTN}.

\begin{figure*}[!t]
    \centering
    \includegraphics[width=\textwidth*10/10, keepaspectratio=true]{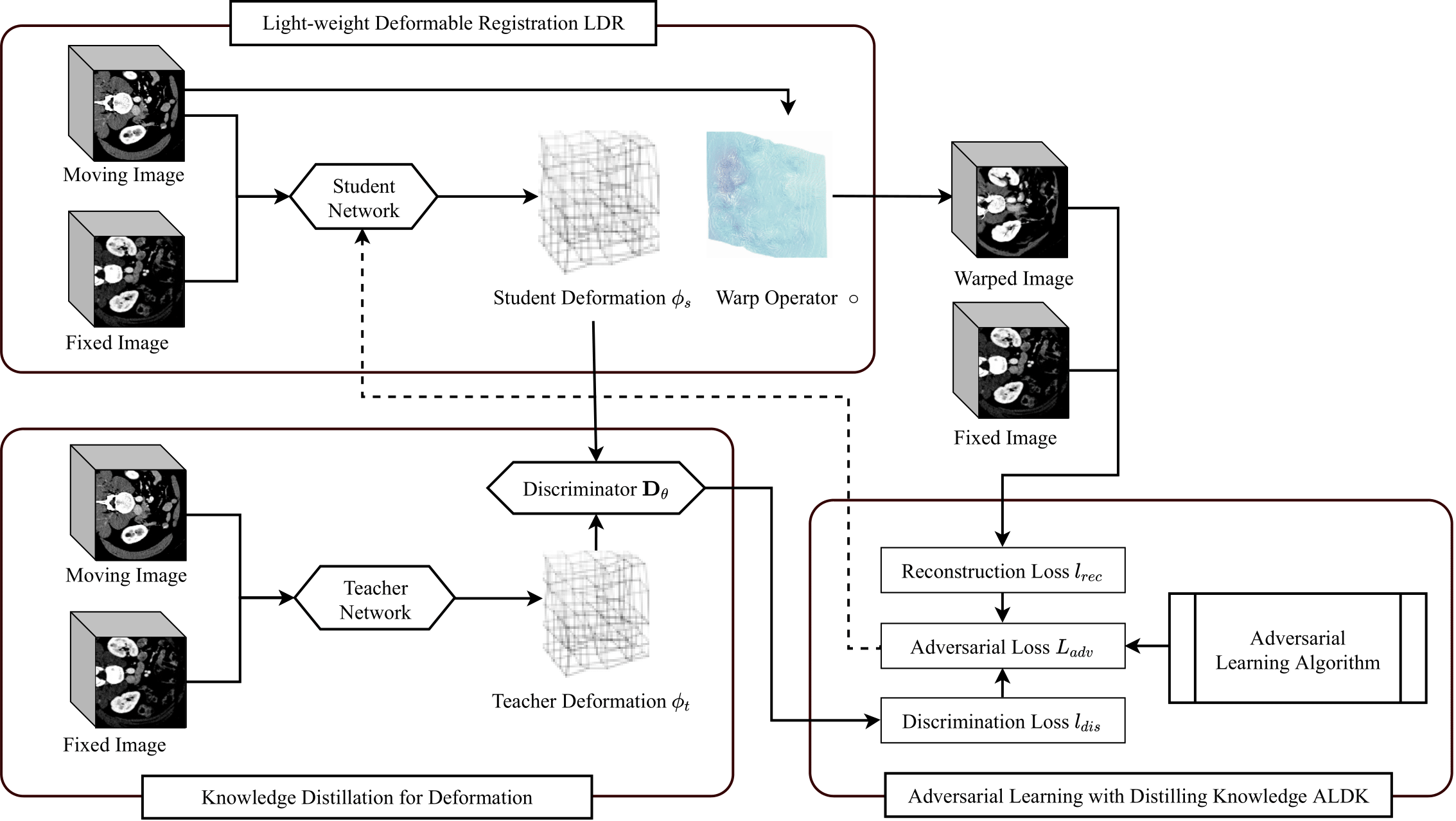}
    \caption{An overview of our proposed Light-weight Deformable Registration (LDR) method using Adversarial Learning with Distilling Knowledge (ALDK). Firstly, by using knowledge distillation, we extract the deformations from the Teacher Network as meaningful ground-truths. Secondly, we design a light-weight student network, which has competitive speed. Finally, We employ the Adversarial Learning with Distilling Knowledge algorithm to effectively transfer the meaningful knowledge of distilled deformations from the Teacher Network to the Student Network.}
    \label{fig:overall_proposal}
\end{figure*}

\textbf{Adversarial Learning}.
Adversarial learning has shown its effectiveness in improving the performance of many registration tasks. Indeed, it works as a powerful regularization method in different supervised generators \cite{nie2020Syn,wgans,yang2018DenseRecontrstruct, Mahapatra2018DeformableMI, fan2018adversarial,chi2020collaborative,yi2019generative, tanner2018generative}. In~\cite{hu2018adversarial}, the authors minimize an additional adversarial generator loss that measures the divergence between the predicted and the biomechanical-based simulated deformations. The authors in~\cite{fan2018adversarial} introduce an unsupervised adversarial similarity network that automatically learns the similarity metric for the deformable registration task without any ground-truth. Recent studies have shown that adversarial data synthesis or augmentation during the training process is effective to improve model generalization and robustness \cite{zhao2018towards,nguyen2020autonomous,fan2019adversarial, yang2020unsupervised,nguyen2019object}.
The use of adversarial learning makes the model output less biased and leverages learnable parameters more efficiently. Different from previous approaches that focus on regularization or augmentation purposes, we take advantage of well-learned deformations extracted from a cumbersome trained model, i.e., distilling knowledge. Thereby, exploring how adversarial learning can be used to improve the performance of our designed light-weight student network.

\textbf{Knowledge Distillation}. 
Knowledge distillation is a process of transferring knowledge from a cumbersome pretrained model (i.e., teacher network) to a smaller light-weight one (i.e., student network) \cite{hinton2015distilling, NguyenDuc2020MEDTEXTA, nayak2019zero, li2020towards}. The light-weight student network is useful in cases where computational resources and deployment costs need to be reduced during the inference stage. For instance, to interpret the teacher model, an explainer module is introduced by \cite{NguyenDuc2020MEDTEXTA} to highlight the regions of an input medical image that are important for the prediction of the teacher model. \hl{In practice, neural networks typically produce class probabilities by using a softmax output layer that converts the logit for each class into a probability by comparing this logit with the other logits}\cite{hinton2015distilling}. In the deformable registration task, inspired by knowledge distillation, we leverage the good pseudo guiding deformations of the teacher model as useful ground-truths in our adversarial learning algorithm to improve the performance of the student network.

Unlike other approaches that mainly target the optimization of the non-linear spatial correspondence between the input images, we introduce a Light-weight Deformable Registration network (LDR) that copes with challenges in model complexity and deployment costs. Our light-weight network is trained using a novel Adversarial Learning with Distilling Knowledge algorithm (ALDK), allowing it to achieve state-of-the-art performance while having fewer parameters and significantly reducing the inference time. 

\section{Methodology}
\label{learning_strat}

\begin{figure*}[!t]
    \centering
    \includegraphics[width=\textwidth*10/10, keepaspectratio=true]{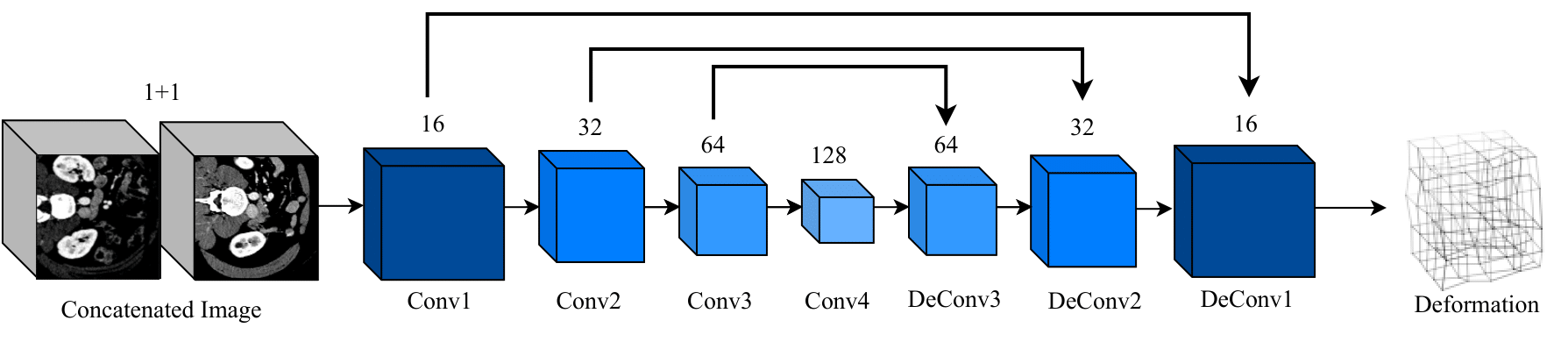}
    \caption{
    The structure of our proposed Light-weight Deformable Registration student network. The number of channels is annotated above the layer. Curved arrows represent skip paths (layers connected by an arrow are concatenated before transposed convolution). Smaller canvas means lower spatial resolution. 
    }
    \label{fig:LightWeight_deformation}
\end{figure*}

\begin{figure*}[!t]
    \centering
    \includegraphics[scale=0.8]{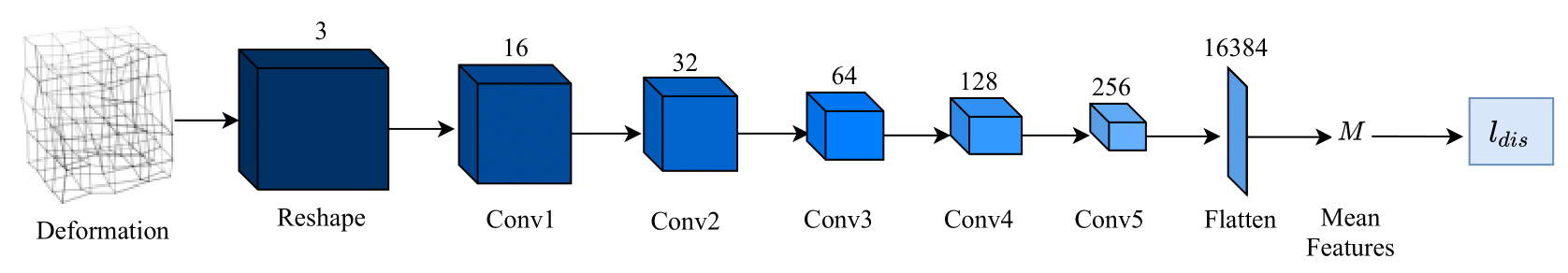}
    \vspace{0.3cm}
    \caption{The structure of the discriminator $D_\mathbf{\theta}$ used in the Discrimination Loss ($l_{dis}$) of our Adversarial Learning with Distilling Knowledge algorithm.
    }
    \vspace{-2ex}
    \label{fig:Ad_CNN}
\end{figure*}

In this section, we describe our method for Light-weight Deformable Registration using Adversarial Learning with Distilling Knowledge. Our method is composed of three main components:
\textit{(i)} a Knowledge Distillation module which extracts meaningful deformations $\bm{\phi_t}$ from the Teacher Network; \textit{(ii)} a Light-weight Deformable Registration (LDR) module which outputs a high-speed Student Network; and \textit{(iii)} an Adversarial Learning with Distilling Knowledge (ALDK) algorithm which effectively leverages teacher deformations $\bm{\phi}_t$ to the student deformations. 
An overview of our proposed deformable registration method can be found in Fig.~\ref{fig:overall_proposal}.

\subsection{Background: Deformable Registration}
\label{subsec:background}
\hl{We follow RCN}\cite{RCN} \hl{to define deformable registration task recursively using multiple cascades}. Let $\textbf{\textit{I}}_m, \textbf{\textit{I}}_ f$ denote the moving image and the fixed image respectively, both defined over $d$-dimensional space $\bm{\Omega}$. A deformation is a mapping $\bm{\phi} : \bm{\Omega} \rightarrow \bm{\Omega}$. A reasonable deformation should be continuously varying and prevented from folding. The deformable registration task is to construct a flow prediction function $\textbf{F}$ which takes $\textbf{\textit{I}}_m, \textbf{\textit{I}}_ f$ as inputs and predicts a dense deformation $\bm{\phi}$ that aligns $\textbf{\textit{I}}_m$ to $\textbf{\textit{I}}_f$ using a warp operator $\circ$ as follows:
\begin{equation}
    \label{eq:deform_compute_ori_in_funct}
    \textbf{F}^{(n)}(\textbf{\textit{I}}^{(n-1)}_m,\textbf{\textit{I}}_f)=\bm{\phi}^{(n)} \circ \textbf{F}^{(n-1)}(\bm{\phi}^{(n-1)} \circ \textbf{\textit{I}}^{(n-2)}_m,\textbf{\textit{I}}_f)
\end{equation}
where $\textbf{F}^{(n-1)}$ is the same as $\textbf{F}^{(n)}$, but in a different flow prediction function. Assuming for $n$ cascades in total, the final output is a composition of all predicted deformations, i.e.,
\begin{equation}
    \label{eq:multi_deformation_compute_ori_in_fuct}
    \textbf{F}(\textbf{\textit{I}}_m, \textbf{\textit{I}}_f)=\bm{\phi}^{(n)} \circ...\circ \bm{\phi}^{(1)},
\end{equation}
and the final warped image is constructed by 
\begin{equation}
    \label{warped_gen}
    \textbf{\textit{I}}_{m}^{(n)}=\textbf{F}(\textbf{\textit{I}}_m,\textbf{\textit{I}}_f) \circ \textbf{\textit{I}}_m
\end{equation}

In general, Equations \ref{eq:deform_compute_ori_in_funct} and \ref{eq:multi_deformation_compute_ori_in_fuct} form the hypothesis function $\mathcal{F}$ under the learnable parameter $\mathbf{W}$,
\begin{equation}
    \label{hypo_funct}
    \mathcal{F}(\textbf{\textit{I}}_{m}, \textbf{\textit{I}}_f, \mathbf{W}) = (\mathbf{v}_{\phi}, \textbf{\textit{I}}_m^{(n)})
\end{equation}
where $\mathbf{v}_{\phi} = [\bm{\phi}^{(1)}, \bm{\phi}^{(2)}, ..., \bm{\phi}^{(k)},..., \bm{\phi}^{(n)}]$ is a vector containing predicted deformations of all cascades. Each deformation $\bm{\phi}^{(k)}$ can  be computed as 
\begin{equation}
    \label{eq:deform_compute}
    \bm{\phi}^{(k)} = {\mathcal{F}}^{(k)}\left(\textbf{\textit{I}}_{m}^{(k-1)}, \textbf{\textit{I}}_f, \mathbf{W}_{\phi^{(k)}}\right)
\end{equation}

To estimate and achieve a good deformation, different networks are introduced to define and optimize the learnable parameter $\mathbf{W}$~\cite{VTN}. 

\subsection{Knowledge Distillation for Deformation}

Knowledge distillation is the process of transferring knowledge from a cumbersome model (teacher model) to a distilled model (student model). The popular way to achieve this goal is to train the student model on a transfer set using a soft target distribution produced by the teacher model. 

Different from the typical knowledge distillation methods that target the output softmax of neural networks as the knowledge~\cite{hinton2015distilling}, in the deformable registration task, we leverage the teacher deformation $\bm{\phi}_t$ as the transferred knowledge. As discussed in \cite{hinton2015distilling}, teacher networks are usually high-performed networks with good accuracy. Therefore, our goal is to leverage the current state-of-the-art Recursive Cascaded Networks (RCN)~\cite{RCN} as the teacher network for extracting meaningful deformations to the student network. The RCN network contains an affine transformation and a large number of dense deformable registration sub-networks designed by VTN~\cite{VTN}. Although the teacher network has expensive computational costs, it is only applied during the training and will not be used during the inference.

\subsection{Light-weight Deformable Registration Network}

In practice, recent deformation networks follow an encoder-decoder architecture and use 3D convolution to progressively down-sample the image, and deconvolution (transposed convolution) to recover spatial resolution~\cite{RCN, VTN}. However, this setup consumes a large number of parameters. 
Therefore, the built models are computationally expensive and time-consuming. To overcome this problem we design a new light-weight student network as illustrated in Fig.~\ref{fig:LightWeight_deformation}. 

In particular, the proposed light-weight network has four convolution layers and three deconvolution layers. Each convolutional layer has a bank of $4 \times 4 \times 4$ filters with strides of $2 \times 2 \times 2$, followed by a ReLU activation function.
The number of output channels of the convolutional layers starts with $16$ at the first layer, doubling at each subsequent layer, and ends up with $128$.
Skip connections between the convolutional layers and the deconvolutional layers are added to help refine the dense prediction. The subnetwork outputs a dense flow prediction field, i.e., a $3$ channels volume feature map with the same size as the input.

In comparison with the current state-of-the-art dense deformable registration network~\cite{VTN}, the number of parameters of our proposed light-weight student network is reduced approximately $10$ times. In practice, this significant reduction may lead to an accuracy drop. Therefore, we propose a new Adversarial Learning with Distilling Knowledge algorithm to effectively leverage the teacher deformations $\bm{\phi}_t$ to our introduced student network, making it light-weight but achieving competitive performance.

\subsection{Adversarial Learning with Distilling Knowledge}
\label{subsec:AD}
Our adversarial learning algorithm aims to improve the student network accuracy through the distilled teacher deformations extracted from the teacher network. The learning method comprises a deformation-based adversarial loss $\mathcal{L}_{adv}$ and its accompanying learning strategy  (Algorithm~\ref{alg:adv_learning}).

\textbf{Adversarial Loss.} The loss function for the light-weight student network is a combination of the discrimination loss $l_{dis}$ and the reconstruction loss $l_{res}$. 
However, the forward and backward process through loss function is controlled by the Algorithm~\ref{alg:adv_learning}.
In particular, the last deformation loss $\mathcal{L}_{adv}$ that outputs the final warped image can be written as:
\begin{equation}
    \label{adv_loss}
    \mathcal{L}_{adv} = \gamma l_{rec} + (1 - \gamma) l_{dis} 
\end{equation}
where $\gamma$ controls the contribution between $l_{rec}$ 
and $l_{dis}$. Note that, the $\mathcal{L}_{adv}$ is only applied for the final warped image.

\textbf{Discrimination Loss.}
In the student network, inspired by WGAN-GP~\cite{wgans}, the discrimination loss is computed in Equation \ref{dis_loss}. 
\begin{equation}
    \label{dis_loss}
    l_{{dis}} = \left\lVert D_\mathbf{\theta}(\bm{\phi}_{s}) - D_\mathbf{\theta}(\bm{\phi}_{t}) \right\lVert_2^{2} + \lambda\bigg(\left\lVert \nabla_{\bm{\hat\phi}_{s}}D_\mathbf{\theta}(\bm{\hat\phi}_{s}) \right\lVert_2 - 1\bigg)^{2}
\end{equation}
where $\lambda$ controls gradient penalty regularization. The joint deformation $\bm{\hat\phi}_{s}$ is computed from the teacher deformation $\bm{\phi}_{t}$ and the predicted student deformation $\bm{\phi}_{s}$ as follow:
\begin{equation}
    \label{joint_deformations}
    \bm{\hat\phi}_{s} = \beta \bm{\phi}_{t} + (1 - \beta) \bm{\phi}_{s}
\end{equation}
where $\beta$ control the effect of the teacher deformation.

In Equation~\ref{dis_loss}, $D_\mathbf{\theta}$ is the discriminator, formed by a neural network with learnable parameters ${\theta}$. The details of $D_\mathbf{\theta}$ is shown in Fig. \ref{fig:Ad_CNN}. In particular, $D_\mathbf{\theta}$ consists of six $3D$ convolutional layers, the first layer is $128 \times 128 \times 128 \times 3$ and takes the $c \times c \times c \times 1$ deformation as input. $c$ is equaled to the scaled size of the input image.
The second layer is $64 \times 64 \times 64 \times 16$. From the second layer to the last convolutional layer, 
each convolutional layer has a bank of $4 \times 4 \times 4$ filters with strides of $2 \times 2 \times 2$, followed by a ReLU activation function except for the last layer which is followed by a sigmoid activation function. 
The number of output channels of the convolutional layers starts with $16$ at the second layer, doubling at each subsequent layer, and ends up with $256$. 
Basically, this is to inject the condition information with a matched tensor dimension and then leave the network learning useful features from the condition input. The output of the last neural layer is the mean feature of the discriminator, denoted as $M$.
Note that in the discrimination loss, a gradient penalty regularization is applied to deal with critic weight clipping which may lead to undesired behavior in training adversarial networks.

\SetKwInput{KwInput}{Input}                
\SetKwInput{KwOutput}{Output}              
\begin{algorithm}[]
\DontPrintSemicolon
  \KwInput{The number of generator iterations per  discriminator iteration $n_{gen}$. The batch size $b$. Initial generator parameters $\mathbf{W}_0$. Initial discriminator parameters $\theta_0$.}
      \While{$(\mathbf{W}$ AND $\theta)$ has not converged}{
      \For{$t = 1$ \KwTo $n_{gen}$}{
        \For{$i = 1$ \KwTo $b$}{
        Sample $x$ real data images in the dataset.
        
        $l^{(i)}_{rec} \leftarrow$ Compute reconstruction loss of $i$-th sample using Equation \ref{res_loss}.
        }
        Backward $l_{rec}$.
        }
        \For{$i = 1$ \KwTo $b$}{
        Sample $x$ real data images in the dataset.
        
        $l^{(i)}_{rec} \leftarrow$ Compute reconstruction loss of $i$-th sample using Equation \ref{res_loss}.
        
        $\bm{\hat\phi}_{s}^{i} \leftarrow$ Compute joint deformation of the final deformation of $i$-th sample using Equation \ref{joint_deformations}.
        
        $l^{(i)}_{dis} \leftarrow$ Compute discrimination loss of the final deformation of $i$-th sample using Equation \ref{dis_loss}.
        
        $\mathcal{L}^{(i)}_{adv} \leftarrow$ Compute adversarial loss of $i$-th sample using Equation \ref{adv_loss}.
        }
        Backward $\mathcal{L}_{adv}$.
      }
\caption{Adversarial Learning Strategy}
\label{alg:adv_learning}
\end{algorithm}

\textbf{Reconstruction Loss.}
The reconstruction loss $l_{rec}$ is an important part of a deformation estimator. Follow the VTN \cite{VTN} baseline, the reconstruction loss is written as:
\begin{equation}
    l_{{rec}}  (\textbf{\textit{I}}_m^h,\textbf{\textit{I}}_f) = 1 - CorrCoef [\textbf{\textit{I}}_m^h,\textbf{\textit{I}}_f]
    \label{res_loss}
\end{equation}
where
\begin{equation}
    CorrCoef[\textbf{\textit{I}}_1, \textbf{\textit{I}}_2] = \frac{Cov[\textbf{\textit{I}}_1,\textbf{\textit{I}}_2]}{\sqrt{Cov[\textbf{\textit{I}}_1,\textbf{\textit{I}}_1]Cov[\textbf{\textit{I}}_2,\textbf{\textit{I}}_2]}}
    \label{coef}
\end{equation}
\begin{equation}
    Cov[\textbf{\textit{I}}_1, \textbf{\textit{I}}_2] = \frac{1}{|\omega|}\sum_{x \in \omega} \textbf{\textit{I}}_1(x)\textbf{\textit{I}}_2(x) -  \frac{1}{|\omega|^{2}}\sum_{x \in \omega} \textbf{\textit{I}}_1(x)\sum_{y \in \omega}\textbf{\textit{I}}_2(y)
    \label{cov}
\end{equation}
where $CorrCoef[\textbf{\textit{I}}_1, \textbf{\textit{I}}_2]$ is the correlation between two images $\textbf{\textit{I}}_1$ and $\textbf{\textit{I}}_2$, $Cov[\textbf{\textit{I}}_1, \textbf{\textit{I}}_2]$ is the covariance between them. $\omega$ denotes the cuboid (or grid) on which the input images are defined.

\textbf{Learning Strategy.}
The forward and backward of the aforementioned $\mathcal{L}_{adv}$ is controlled by the adversarial learning strategy described in Algorithm \ref{alg:adv_learning}.

In our deformable registration setup, the role of real data and attacking data is reversed when compared with the traditional adversarial learning strategy. \hl{In adversarial learning}~\cite{wgans}, \hl{the model uses unreal (generated) images as attacking data, while image labels are ground truths. However, in our deformable registration task, the model leverages the unreal (generated) deformations from the teacher as attacking data, while the image is the ground truth for the model to reconstruct the input information. As a consequence, the role of images and the labels are reversed in our setup.} Since we want the information to be learned more from real data, the generator will need to be considered more frequently. Although the knowledge in the discriminator is used as attacking data, the information it supports is meaningful because the distilled information is inherited from the high-performed teacher model. With these characteristics of both the generator and discriminator, the light-weight student network is expected to learn more effectively and efficiently.

\section{Experiments}
\label{sec_exps}
\subsection{Datasets}
\label{datasets}
We generally follow~\cite{RCN, VTN} to conduct our experiments. In particular, we train our method on two types of scans: Liver CT scans and Brain MRI scans.

\begin{table*}[!t]
\centering
\setlength{\tabcolsep}{1.1 em} 
{\renewcommand{\arraystretch}{1.4}
\begin{tabular}{|l|c|c|c|c|c|c|c|c|c|c|c|}
\hline
\multicolumn{1}{|c|}{\multirow{3}{*}{\textbf{Architecture}}} & \multicolumn{2}{c|}{\textbf{SLIVER}} & \multicolumn{2}{c|}{\textbf{LiST}} & 
\multicolumn{2}{c|}{\textbf{LSPIG}} & \multicolumn{2}{c|}{\textbf{LPBA}}                      &
\multicolumn{1}{c|}{\multirow{3}{*}{\textbf{\begin{tabular}[c]{@{}c@{}}CPU\\ (sec)\end{tabular}}}}&
\multicolumn{1}{c|}{\multirow{3}{*}{\textbf{\begin{tabular}[c]{@{}c@{}}GPU\\ (sec)\end{tabular}}}}&
\multirow{3}{*}{\textbf{\begin{tabular}[c]{@{}c@{}}\#Params\end{tabular}}}\\ \cline{2-9}
\multicolumn{1}{|c|}{}                                       & \textbf{Dice}    & \textbf{Jacc}    & \textbf{Dice}    & \textbf{Jacc}   & \textbf{Dice}    & \textbf{Jacc} &\textbf{Dice}    & \textbf{Jacc}    & & &                    \\ 
\multicolumn{1}{|c|}{}                                       & ($\%$)    & ($\%$)    & ($\%$)    & ($\%$)   & ($\%$)    & ($\%$) &($\%$)    & ($\%$)    & &   &                 
\\ \hline
ANTs SyN~\cite{ANTs}                                             & $89.5$                 & $81.2$                 & $86.2$      &   $\--$        & $82.5$      &   $\--$                        & $70.8$                 & $\--$                  & $748$  & $\--$  & $\--$                                                                            \\ \hline
Elastix B-spline~\cite{Elastix}                                             & $91.0$                 & $83.7$                  & $86.3$       &  $\--$           & $82.5$      &   $\--$               & $67.5$                    &  $\--$              & $115$     & $\--$     &   $\--$                                                                     \\ \hline
VoxelMorph~\cite{VoxelMorph}                                       & $91.3$                  & $84.0$                 & $87.0$         & $76.2$          & $83.3$      &   $\--$                        & $68.8$            &  $\--$                      & $14$  & $0.31$ & $14.47$M                                                                            \\ \hline
VTN (ADDD)~\cite{VTN}                                                   & $94.2$                  & $88.6$                  & $89.7$         & $\--$           & $84.6$      &   $\--$                     & $70.1$              & $\--$                     & $26$      & $0.28$  & $98.90$M                                                                       \\ \hline
1-cas \hl{RCN}~\cite{RCN}                                              & $91.4$                 & $86.1$                  & $87.0$        & $78.4$         & $83.3$      &   $72.8$                       & $68.6$               &  $53.4$                  & $10$     & $0.20$   & $42.41$M                                                                      \\ \hline
2-cas \hl{RCN}~\cite{RCN}                                              & $93.5$                 & $87.9$                  & $89.1$        & $80.5$         & $84.3$      &   $73.7$                       & $69.7$               &  $54.2$                  & $18$     & $0.31$   & $70.67$M                                                                      \\ \hline

3-cas \hl{RCN}~\cite{RCN}                                              & $94.3$                 & $89.3$                  & $90.0$           & $82.7$   & $85.0$      &   $74.4$                            &  $70.3$              & $55.0$                   & $26$    & $0.40$    & $98.89$M                                                                       \\ \hline 
1-cas LDR + ALDK (ours)                                        & $91.2$                 & $83.6$               & $86.7$        & $76.4$    & $83.7$ & $72.3$ &$68.3$ & $53.0$ & $0.69$                                                                          &   $0.16$    &$0.56$M                                                                   \\ \hline

2-cas LDR + ALDK (ours)                                      & $93.2$                 & $86.0$               & $88.6$        & $78.7$   & $84.1$ & $73.2$ & $69.0$ & $53.5$ & $0.98$                                                                        &  $0.24$       &$0.84$M                                                                    \\ \hline
3-cas LDR + ALDK (ours)                                       & $94.0$                 & $87.1$                   & $89.4$              &  $81.1$           & $84.6$      &   $73.9$                             & $69.6$       & $54.3$                & $1.24$      &    $0.30$ & $1.12$M                                                                     \\ \hline
\end{tabular}
}
\vspace{1ex}
\caption{Comparison among our proposed model with recent approaches.
}
\label{tab:comparison}
\end{table*}

For Liver CT scans, we use $5$ datasets: 

\begin{enumerate}
    \item LiTS \cite{LiTS} contains 131 liver segmentation scans. 
    \item MSD \cite{MSD} has 70 liver tumor CT scans, 443 hepatic vessels scans, and 420 pancreatic tumor scans.
    \item BFH \cite{VTN} is a smaller dataset with 92 scans.
    \item SLIVER \cite{heimann2009comparison} is a challenging dataset with 20 liver segmentation scans and annotated by $3$ expert doctors.
    \item LSPIG (Liver Segmentation of Pigs) contains 17 pairs of CT scans from pigs, provided by the First Affiliated Hospital of Harbin Medical University.
\end{enumerate}
For Liver CT scans, all methods are trained on the combination of MSD and BFH datasets with $1025^2$ $(1025 = 70 + 443 + 420 + 92)$ image pairs in total. The SLIVER $(20 \times 19$ pairs$)$, LiTS $(131 \times 130$ pairs$) $, and LSPIG $(34$ intra-subject pairs$)$ datasets are used for evaluation. 

For Brain MRI scans, we use $4$ datasets: 
\begin{enumerate}
    \item ADNI \cite{mueller2005ways} contains $66$ scans.
    \item ABIDE \cite{di2014autism} contains $1287$ scans.
    \item ADHD \cite{bellec2017neuro} contains $949$ scans.
    \item LPBA \cite{shattuck2008construction} has $40$ scans, each featuring a segmentation ground truth of $56$ anatomical structures.
\end{enumerate}
For Brain MRI scans, as in~\cite{RCN,VTN}, the ADNI, ABIDE, and ADHD dataset are used for training, and the LPBA dataset is used for testing. 

\subsection{Experimental Setup}
\label{metrics}
\textbf{Evaluation Metric.} As standard practice~\cite{RCN}, we use the Dice score to quantify the performance of all models. 
The Dice score can be computed as: 
\begin{equation}
    \label{dice_score}
    Dice(A,B) = 2 \cdot \frac{|A \cap B|}{|A| + |B| }
\end{equation}

Also, the Jaccard correlation coefficient (Jacc) 
between the warped segmentation and the ground-truth can be utilized as an auxiliary metric~\cite{VTN}.
\begin{equation}
    \label{jac_score}
    Jacc(A,B) = \left|\frac{A \cap B}{A \cup B }\right|
\end{equation}
where $A, B$ are the set of voxels the organ consists of.

To verify the effectiveness of all models in practice, we also report the performance in terms of speed (CPU and GPU time - second/sample) and the number of network parameters during the inference stage.

\textbf{Baseline.} We compare our proposed method with the following recent deformable registration methods:
\begin{itemize}
    \item ANTs SyN~\cite{ANTs} and Elastix B-spline~\cite{Elastix} are methods that find an optimal transformation by iteratively update the parameters of the defined alignment.
    \item VoxelMorph~\cite{VoxelMorph} predicts a dense deformation in an unsupervised manner by using deconvolutional layers.
    \item VTN \cite{VTN} is an end-to-end learning framework that uses convolutional neural networks to register 3D medical images, especially large displaced ones.
    \item RCN \cite{RCN} is a recent recursive deep architecture that utilizes learnable cascade and performs progressive deformation for each warped image.
\end{itemize}
\hl{It is worth noting that RCN is cascaded VTN. Therefore, both RCN and VTN can be considered as state-of-the-art approaches. In practice, we keep the architecture of all teacher networks unchanged.}

\textbf{Implementation.} 
We implement our network using TensorFlow~\cite{singh2020introductionTF}. The network is trained with a batch size of $4$ on $11$GB Nvidia 1080 Ti. The training stage runs for $105$ iterations and takes approximately $8$ hours with Adam optimizer~\cite{kingma2014adam}. 
The learning rate is set to $10^{-4}$. Based on validation results, the parameters $n_{gen}$, $\beta$, $\lambda$, and $\gamma$ are set to $3$, $0.1$, $1.0$, and $0.5$ for CT scans and $3$, $0.05$, $0.9$, and $0.3$ for MRI scans, respectively. 

\textbf{Hardware Setup.} Since measuring the inference time is crucial to compare the effectiveness of all methods in practice,  we test and report the inference time of all baselines and our method on the same CPU and GPU. They are Intel Xeon E5-2690 v4 CPU and Nvidia GeForce GTX 1080 Ti GPU, respectively. No overclocking is used.

\subsection{Results}

Table~\ref{tab:comparison} summarizes our overall performance, testing speed, and the number of parameters compared with recent state-of-the-art methods in the deformable registration task. The results clearly show that our proposed Light-weight Deformable Registration network (\text{LDR}) accompanied by our Adversarial Learning with Distilling Knowledge (\text{ALDK}) algorithm significantly reduces the inference time and the number of parameters during the inference phase. Moreover,  the proposed method achieves competitive accuracy with the most recent highly performed but expensive networks, such as VTN or VoxelMorph. We notice that this improvement is consistent across all experiments on different datasets SLIVER, LiST, LSPIG, and LPBA.

\begin{table*}[!t]
\centering
\setlength{\tabcolsep}{0.4 em} 
{\renewcommand{\arraystretch}{1.4}
\begin{tabular}{|l|c|c|c|c|c|c|c|c|c|c|}
\hline
\multicolumn{1}{|c|}{\multirow{3}{*}{\textbf{Architecture}}} & \multicolumn{2}{c|}{\textbf{SLIVER}} & \multicolumn{2}{c|}{\textbf{LiST}} & 
\multicolumn{2}{c|}{\textbf{LSPIG}} & \multicolumn{2}{c|}{\textbf{LPBA}}                      &
\multicolumn{1}{c|}{\multirow{3}{*}{\textbf{\begin{tabular}[c]{@{}c@{}}CPU\\ (sec)\end{tabular}}}}&
\multirow{3}{*}{\textbf{\begin{tabular}[c]{@{}c@{}}\#Params\end{tabular}}}\\ \cline{2-9}
\multicolumn{1}{|c|}{}                                       & \textbf{Dice}    & \textbf{Jacc}    & \textbf{Dice}    & \textbf{Jacc}   & \textbf{Dice}    & \textbf{Jacc} &\textbf{Dice}    & \textbf{Jacc}    & &                    \\ 
\multicolumn{1}{|c|}{}                                       & ($\%$)    & ($\%$)    & ($\%$)    & ($\%$)   & ($\%$)    & ($\%$) &($\%$)    & ($\%$)    & &                    
\\ \hline
1-cas LDR                   & $87.8$                 & $78.4$                 & $82.9$        & $71.2$     &$75.9$ &$67.4$ &$66.8$ &$51.4$ &$0.69$                                                                           &$0.56$M                                                                               \\

1-cas LDR + ALDK                                        & $91.2 (+3.4)$                 & $83.6 (+5.2)$               & $86.7 (+3.8)$        & $76.4(+5.2)$    & $83.7 (+7.8)$ & $72.3 (+4.9)$ &$68.3 (+1.4)$ & $53.0 (+1.6)$ & $0.69$                                                                               &$0.56$M                                                                   \\ \hline

2-cas LDR                   & $89.2$                 & $81.1$                 & $85.5$        & $74.2$  &$78.6$ &$69.9$ &$67.9$ &$52.2$    &    $0.98$                                                                           &$0.84$M                                                                               \\
2-cas LDR + ALDK                                    & $93.2 (+4.0)$                 & $86.0 (+4.9) $               & $88.6 (+3.1)$        & $78.7 (+4.5)$   & $84.1 (5+5.5)$ & $73.2 (+3.3)$ & $69.0 (+1.1)$ & $53.5(+1.3)$ & $0.98 $                                                                               &$0.84$M                                                                    \\ \hline

3-cas LDR                   & $90.9$                 & $83.2$                 & $87.9$        & $77.6$   & $81.2$ &$71.5$ &$68.5$ &$52.7$    & $1.24$        & $1.12$M                                                                              \\

3-cas LDR + ALDK                                      & $94.0 (+3.1)$                 & $87.1 (+3.9)$                   & $89.4 (+1.5)$              &  $81.1 (+3.5)$           & $84.6 (+3.4)$      &   $73.9 (+2.4)$                             & $69.6 (+1.1)$       & $54.3 (+1.6)$                & $1.24$        & $1.12$M                                                                     \\ \hline
\end{tabular}
}
\vspace{1ex}
\caption{The comparison when our Light-weight Deformable Registration network is used with and without the  Adversarial Learning with Distilling Knowledge procedure.}
\vspace{-3ex}
\label{tab:abl_lightweight}
\end{table*}

In particular, we observe that on the SLIVER dataset the Dice score of our best model with $3$ cascades (3-cas LDR + ALDK) is $0.3\%$ less than the best result of 3-cas VTN + Affine, while our inference speed is $\sim21$ times faster on a CPU and the parameters used during inference is $\sim88$ times smaller. Including benchmarking results in three other datasets, i.e., LiST, LSPIG, and LPBA, our proposed light-weight model only trades off an average of $0.5\%$ in Dice score and $1.25\%$ in Jacc score for a significant gain of speed and a massive reduction in the number of parameters. We also notice that our method is the only work that achieves the inference time of approximately $1s$ on a CPU. This makes our method well suitable for deployment as it does not require expensive GPU hardware for inference.

\subsection{Ablation Study}
\label{sec:abl}

\textbf{Effectiveness of ALDK.}
Table \ref{tab:abl_lightweight} summarizes the effectiveness of our proposed Adversarial Learning with Distilling Knowledge (ALDK) when being integrated into the light-weight student network. \hl{Note that LDR without ALDK is trained using only the reconstruction loss in an unsupervised learning setup.} From this table, we clearly see that our proposed ALDK algorithm improves the Dice score of the LDR tested in the SLIVER dataset by $+3.4\%$, $+4.0\%$, and $+3.1\%$ for 1-cas, 2-cas, and 3-cas setups, respectively. Additionally, using ALDK also increases the Jacc score by $+5.2\%$, $+4.9\%$, and $3.9\%$ for 1-cas LDR, 2-cas LDR, and 3-cas LDR. These results verify the stability of our adversarial learning algorithm in the inference phase, under the differences evaluation metrics, as well as the number of cascades setups. Furthermore, Table \ref{tab:abl_lightweight} also clearly shows the effectiveness and generalization of our ALDK when being applied to the student network.
Since the deformations extracted from the teacher are used only in the training period, our adversarial learning algorithm fully maintains the speed and the number of parameters for the light-weight student network during inference. 
All results indicate that our student network incorporated with the adversarial learning algorithm successfully achieves the performance goal, while maintaining the efficient computational cost of the light-weight setup. 

\begin{figure*}[]
\centering
    \subfigure[]{\includegraphics[width=0.48\textwidth]{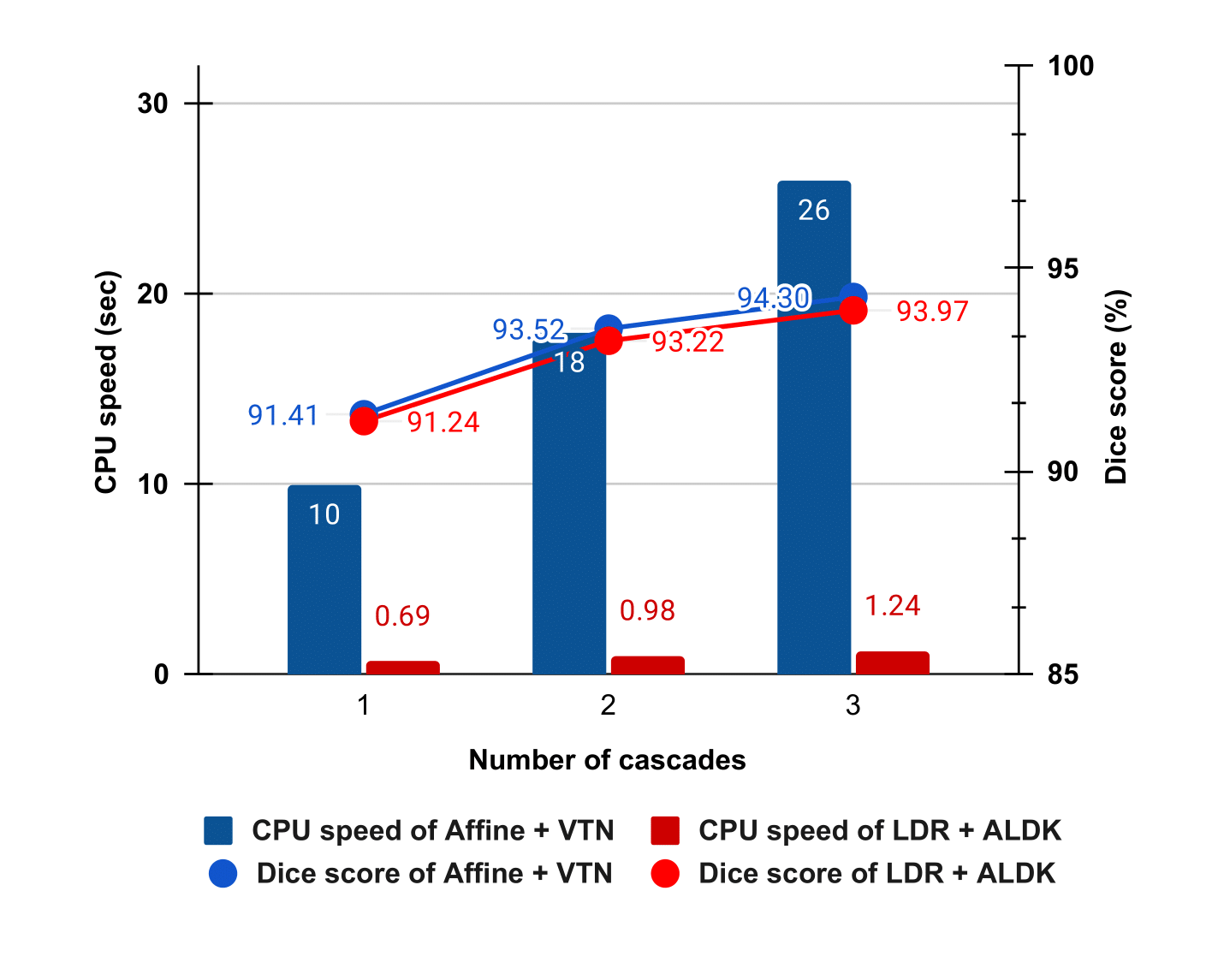}}
    \subfigure[]{\includegraphics[width=0.48\textwidth]{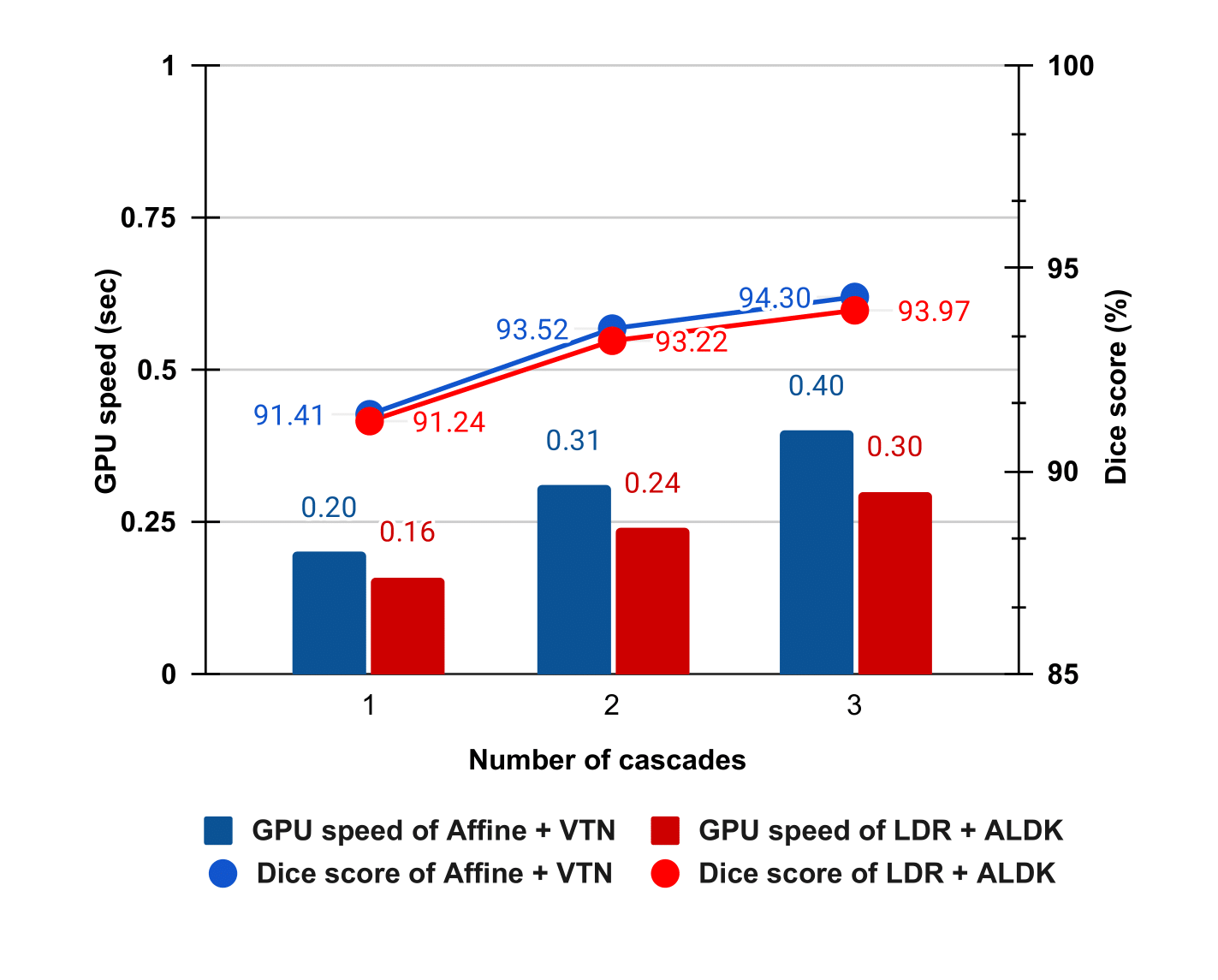}} 
\caption{Plots of Dice score and Inference speed with respect to the number of cascades of the baseline Affine + VTN and our proposed LDR + ALDK. \textbf{(a)} for CPU speed and \textbf{(b)} for GPU speed. Note that results are reported for the SLIVER dataset; bars represent the CPU speed; lines represent the Dice score. All methods use an  Intel Xeon  E5-2690 v4 CPU and Nvidia GeForce GTX 1080 Ti GPU for inference.}
\vspace{-2ex}
\label{fig_vis_graph}
\end{figure*}

\textbf{Accuracy vs. Complexity.} 
Fig. \ref{fig_vis_graph} demonstrates the experimental results from the SLIVER dataset between our LDR + ALDK and the baseline VTN \cite{VTN} under multiple recursive cascades setup on both CPU and GPU. On the CPU (Fig. \ref{fig_vis_graph}-a), in terms of the 1-cascade setup, the Dice score of our method is $0.2\%$ less than VTN while the speed is $\sim 15$ times faster. The more the number of cascades is leveraged, the higher the speed gap between our proposed LDR + ALDK and the baseline VTN, e.g. the CPU speed gap is increased to $\sim 21$ times in a 3-cascades setup. We also observe the same effect on GPU (Fig. \ref{fig_vis_graph}-b), where our method achieves slightly lower accuracy results than VTN, while clearly reducing the inference time.
These results indicate that our proposed LDR + ALDK can work well with the teacher network to improve the accuracy while significantly reducing the inference time on both CPU and GPU in comparison with the baseline VTN network. 

\subsection{Visualization} 
\begin{figure*}[!t]
  \centering
    \subfigure[LDR]{\includegraphics[width=0.13\linewidth, height=0.67\linewidth]{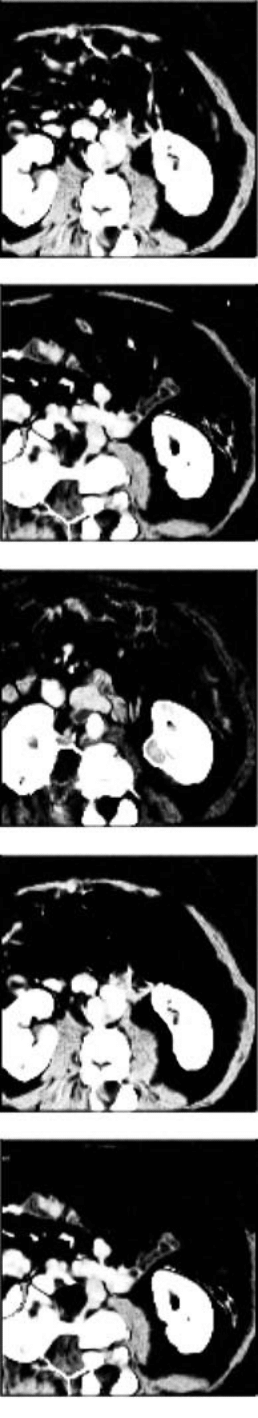} \includegraphics[width=0.13\linewidth, height=0.67\linewidth]{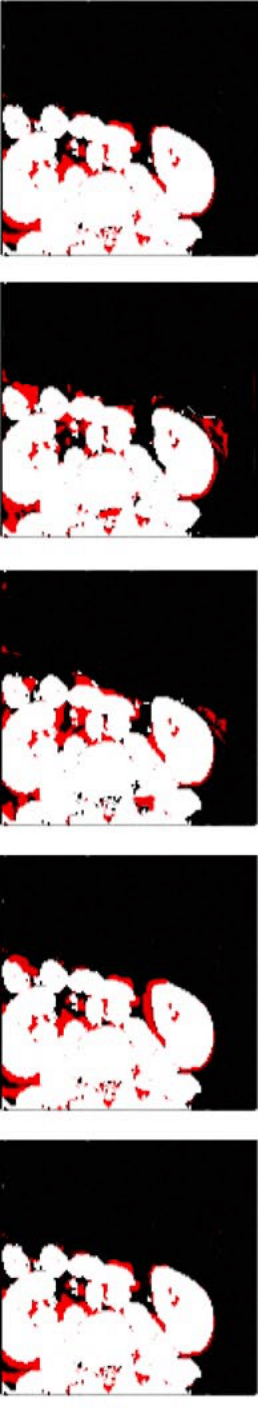}}\hspace{2ex}
    \subfigure[LDR + ALDK (ours)]{\includegraphics[width=0.13\linewidth, height=0.67\linewidth]{images/LDRALDKWarp.png}
    \includegraphics[width=0.13\linewidth, height=0.67\linewidth]{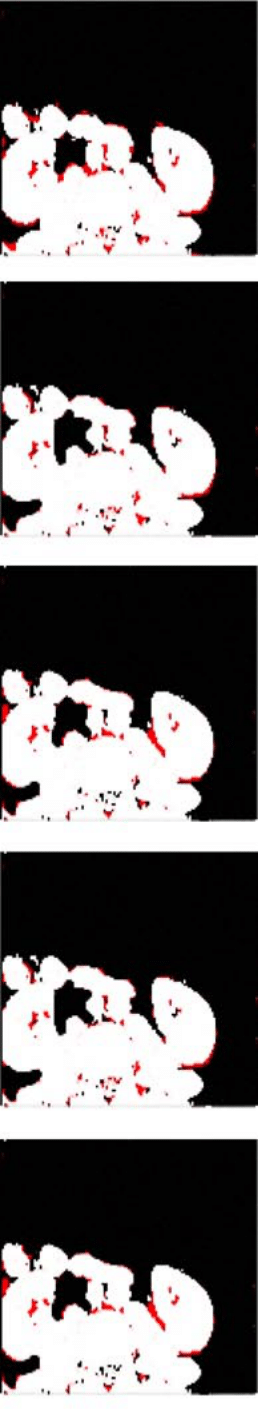}}\hspace{2ex}
    \subfigure[RCN]{\includegraphics[width=0.13\linewidth, height=0.67\linewidth]{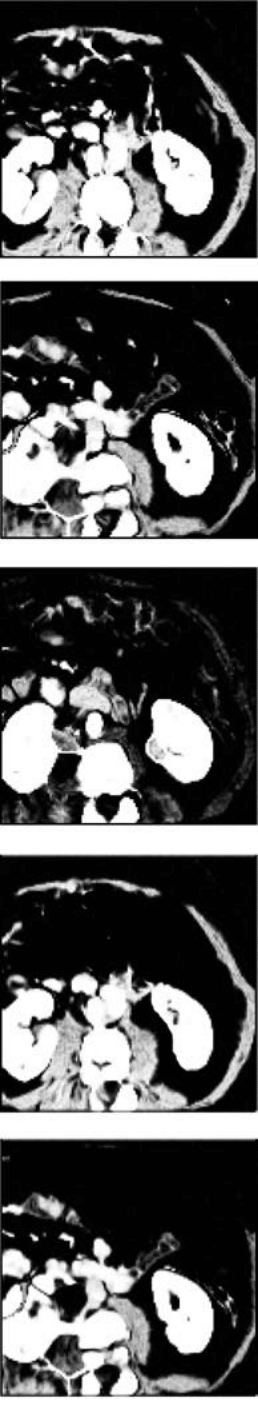}
    \includegraphics[width=0.13\linewidth, height=0.67\linewidth]{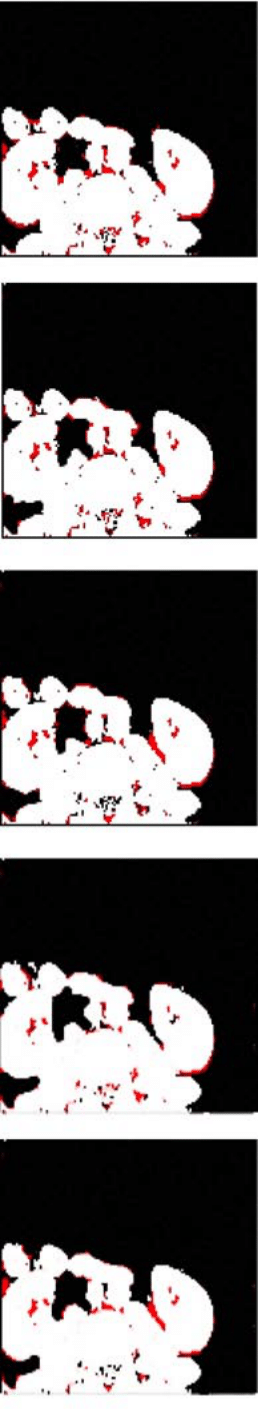}}
 \caption{\hl{The visualization comparison between our LDR (\textbf{a}), LDR + ALDK (\textbf{b}), and the baseline RCN (\textbf{c}).
The left images are sections of the warped images; the right images are sections of the warped segmentation (white color represents the matched areas between warped image and fixed image, red color denotes the miss-matched areas). The segmentation visualization indicates that our proposed LDR + ALDK (\textbf{b}) method reduces the miss-matched areas of the student network LDR (\textbf{a}) significantly. Best viewed in color.}}
 \label{fig:Deformation_Vis}
 \vspace{-2ex}
\end{figure*}

Fig. \ref{fig:Deformation_Vis} illustrates the visual comparison among 1-cas LDR, 1-cas LDR + ALDK, and the baseline 1-cas \hl{RCN}. Five different moving images in a volume are selected to apply the registration to a chosen fixed image. It is important to note that though the sections of the warped segmentations can be less overlap with those of the fixed one, the segmentation intersection over union is computed for the volume and not the sections.
In the segmented images in Fig.~\ref{fig:Deformation_Vis}, besides the matched area colored by white, we also marked the miss-matched areas by red for an easy-to-read purpose.

From Fig. \ref{fig:Deformation_Vis}, \hl{we can see that the segmentation resutls of 1-cas LDR network without using ALDK} (Fig.~\ref{fig:Deformation_Vis}\hl{-a}) contains many miss-matched areas (denoted in red color). However, when we apply ALDK to the student network, the registration results are clearly improved (Fig.~\ref{fig:Deformation_Vis}\hl{-b}). Overall, our LDR + ALDK visualization results in Fig.~\ref{fig:Deformation_Vis}\hl{-b} are competitive with the baseline \hl{RCN} network (Fig.~\ref{fig:Deformation_Vis}\hl{-c}). This visualization confirms that our proposed framework for deformable registration can achieve comparable results with the recent \hl{RCN} network.

\subsection{Student Networks Comparison}
\begin{table}[!ht]
\centering
\setlength{\tabcolsep}{0.4 em} 
{\renewcommand{\arraystretch}{1.5}
\begin{tabular}{|l|c|c|c|c|c|c|c|}
\hline
\multicolumn{1}{|c|}{\multirow{3}{*}{\textbf{\hl{Student Network}}}} & \multicolumn{2}{c|}{\textbf{SLIVER}} & \multicolumn{2}{c|}{\textbf{LiST}} & 
\multicolumn{1}{c|}{\multirow{3}{*}{\textbf{\begin{tabular}[c]{@{}c@{}}CPU\\ (sec)\end{tabular}}}}&
\multicolumn{1}{c|}{\multirow{3}{*}{\textbf{\begin{tabular}[c]{@{}c@{}}GPU\\ (sec)\end{tabular}}}}&
\multirow{3}{*}{\textbf{\begin{tabular}[c]{@{}c@{}}\#Params\end{tabular}}}\\ \cline{2-5}
\multicolumn{1}{|c|}{}                                       & \textbf{Dice}    & \textbf{Jacc}    & \textbf{Dice}    & \textbf{Jacc}     & &   &                 \\ 
\multicolumn{1}{|c|}{}                                       & ($\%$)    & ($\%$)    & ($\%$)    & ($\%$)     &   &  &               
\\ \hline
LDR (ours)                               &$91.2$ &$83.6$ &$86.7$ &$76.4$ &$0.69$ &$0.16$ &$0.56$M     \\ \hline
Mobilenet \cite{howard2017mobilenets}                              &$79.4$ &$75.9$ &$80.0$ &$72.1$      &$4$   &$0.17$    &$1.25$M          \\ \hline
U-Net \cite{ronneberger2015UNet}                               &$86.2$ &$81.8$ &$84.2$ &$75.5$          &$11$   &$0.18$    &$28.28$M      \\ \hline
VoxelMorph \cite{VoxelMorph}                             &$91.6$ &$84.5$ &$87.2$ &$76.4$    &$14$ &$0.31$ &$14.47$M                \\ \hline
VTN \cite{VTN}                            &$89.5$ &$81.6$ &$85.9$ &$76.1$      &$8$   &$0.18$  &$28.25$M   \\\hline
RCN\cite{RCN}                           &    $91.7$              & $86.4$                  & $87.1$     & $78.6$      &$10$  &$0.20$  &$42.41$M           \\\hline
\end{tabular}
}
\caption{The effectiveness of different network architectures when we use them as the student network under our Adversarial Learning with Distilling Knowledge algorithm.}
\vspace{-2ex}
\label{tab:ad_baseline}
\end{table}
The student network plays an important role in our approach as it receives the transferred knowledge from the teacher network during the adversarial learning process. In practice, the student network must also be light-weight to have a fast inference time while being able to maintain the accuracy from the teacher network. Our student LDR network is designed based on the popular encoder-decoder U-Net~\cite{ronneberger2015UNet} architecture. In this study, we show the comparison between different network architectures when we use them as the student network. In particular, we compare our LDR network with the following works: the classical U-Net \cite{ronneberger2015UNet}, the popular light-weight network for computer vision task, i.e., Mobilenet \cite{howard2017mobilenets}, and different networks in recent deformation registration tasks: VoxelMorph \cite{VoxelMorph}, VTN \cite{VTN}, and RCN\cite{RCN}.

Table \ref{tab:ad_baseline} shows the effectiveness of our introduced light-weight student network in comparison with different other architectures when we use them as the student network under ALDK.  Particularly, our LDR achieves comparative results with other high-complexity architectures such as VoxelMorph and RCN. Moreover, our LDR also outperforms U-Net and VTN by a fair margin. We notice that while the accuracy of our LDR is competitive or better than other networks, the inference time and the number of parameters in our LDR network is significantly lower. It is also worth noting that, compared to Mobilenet, which is a well-known light-weight architecture for computer vision tasks, our LDR network outperforms Mobilenet in both accuracy and running time by a large margin. These results indicate that our LDR is a portable architecture, and it works well with ALDK to reserve the accuracy of the teacher network.

\hl{\textbf{Correlation between student and teacher architecture.}
Basically, our proposed ALDK can leverage the deformations extracted from any high-performed teacher model that can extract distilled deformations to improve the student score. However, according to} \cite{cho2019efficacy,liu2020search,han2021fixing}\hl{, the knowledge distillation-based algorithm is more effectively utilized when both the student and teacher have similarities in learning behavior. As shown in Table}~\ref{tab:ad_baseline}\hl{, VTN} \cite{VTN}\hl{, VoxelMorph} \cite{VoxelMorph}\hl{, RCN} \cite{RCN} \hl{achieve good results when we use them as student models since their architectures are very similar to the RCN teacher model}. \hl{From Table} \ref{tab:ad_baseline}\hl{, we can see that MobileNet}\cite{howard2017mobilenets} \hl{achieves less improvement than other students when ALDK is applied since its architecture does not fit well with the teacher model.}

\hl{\textbf{Student Network Complexity.} Through the experimental results, we found out that apart from the correlation between student and teacher model, the complexity of the student network also plays an essential role in improving the results. Since we use adversarial training to transfer the knowledge from the teacher model to the student network, using some cumbersome network may cause difficulty in this process (in adversarial learning, bigger networks may not always converge well because the training optimizes the equilibrium between generator and discriminator loss} \cite{wgans})\hl{. As a result, although U-Net and our LDR share encoder-decoder architecture, U-Net is much bigger than our LDR, hence it is more challenging for U-Net to learn the deformation from the teacher model through the adversarial training process.}

\subsection{Teacher Networks Analysis}

\begin{table*}[!ht]
\centering
\setlength{\tabcolsep}{0.5 em} 
{\renewcommand{\arraystretch}{1.8}
\begin{tabular}{|c|c|c|c|c|c|c|c|}
\hline
\multirow{2}{*}{{\textbf{Student Network}}} & \multirow{2}{*}{{\textbf{Teacher Model}}} & \multicolumn{3}{c|}{{\textbf{SLIVER}}}                                  & \multicolumn{3}{c|}{{\textbf{LiST}}}                                                                     \\ \cline{3-8} 
                                          &                                         & {\textit{Dice (\%)}} & {\textit{Jacc (\%)}} & {\textit{$|J_{\phi}|\leq 0$}} & {\textit{Dice (\%)}} & {\textit{Jacc (\%)}} & {\textit{$|J_{\phi}|\leq 0$}} \\ \hline
{1-cas LDR}                                 & {None}                                    & {87.8}               & {78.4}               & {59,025}                      & {82.9}               & {71.2}               & {39,870}                                                       \\ \hline
{1-cas LDR + ALDK}                                 & {VoxelMorph}~\cite{VoxelMorph}                             & {88.6 (+0.8)}        & {80.2 (+1.8)}        & {54,271}                      & {84.4 (+1.5)}        & {73.7 (+2.5)}        & {37,372}                                                       \\ \hline
{1-cas LDR + ALDK}                                 &     {VTN (ADDD)}~\cite{VTN}                                    & {90.1 (+2.3)}        & {81.8 (+3.4)}        & {48,284}                      & {85.3 (+2.4)}        & {74.8 (+3.6)}        & {36,421}                                                       \\ \hline
{1-cas LDR + ALDK}                                 & {RCN}~\cite{RCN}                                     & {91.2 (+3.4)}        & {83.6 (+5.2)}        & {43,625}                      & {86.7 (+3.8)}        & {76.4 (+5.2)}        & {24,903}                                                       \\ \hline
\end{tabular}
}
\caption{\hl{The effectiveness of different network architectures when we use them as the teacher network under our Adversarial Learning with Distilling Knowledge algorithm.}}
\vspace{-2ex}
\label{tab:teacher_comparison}
\end{table*}

\hl{Table} \ref{tab:teacher_comparison} \hl{illustrates the effectiveness of different models when we use them as the teacher network. We use VoxelMorph} \cite{VoxelMorph}\hl{, ADDD} \cite{VTN}\hl{, and RCN }\cite{RCN}\hl{ as the teacher model since they have high performance in deformable registration. The 1-cas LDR is used as the student network. We also show the mean of the number of voxels, with non-positive Jacobian determinants for all flows in each dataset. From Table} \ref{tab:teacher_comparison}\hl{, we can see that our LDR student network can work well with different teacher models. We achieve a consistent improvement on both SLIVER and LiST datasets. This table also confirms that the use of teacher model and our ALDK is essential to improve the accuracy of the deformation registration task. }

\hl{To further evaluate the results, as in VoxelMorph} \cite{VoxelMorph}, \hl{we analyse the smoothness of deformations extracted from a model using the distortion of the deformation.} \hl{Table }\ref{tab:teacher_comparison}\hl{ also shows the quality of the deformation based on the smoothness calculated by the non-positive Jacobian determinant $J_{\phi}$. The smoothness of deformation is defined by the distortion which is computed by the number of voxels in deformation with non-positive Jacobian determinants. The smaller $J_{\phi}$ is, the better the model as it has less distortion.
Table }\ref{tab:teacher_comparison}\hl{ shows that the smoothness of the deformation using our LDR is improved gradually when the teacher contains the smooth regularizer (VoxelMorph) or has invertibility (VTN, RCN). This confirms that if the teacher contains deformation smoothing techniques to deal with the distortion problem, then our student network can inherit them during the adversarial learning process.}

\section{Discussion and Conclusion}
\label{sec_con}
We introduce a Light-weight Deformable Registration (LDR) network that significantly reduces the model complexity while achieving competitive accuracy. We show that by combining our LDR with the proposed Adversarial Learning with Distilling Knowledge (ALDK) algorithm, our framework can effectively leverage the knowledge of the effective but computationally expensive teacher network to the student network, hence ensuring the student network is light-weight and novel. Currently, our deformable registration framework relies on the teacher model to provide the master knowledge. Although the teacher model is complex, it is only be used during the training process and does not affect the inference time of the light-weight student network.

The extensive experiments confirm that our proposed LDR with ALDK successfully balances the trade-off between the computational costs and model accuracy for the deformable registration task. In the future, we would like to improve ALDK by utilizing more than one teacher network. Furthermore, investigating new light-weight student architectures that can achieve real-time speed on the CPU is also an interesting research direction. Finally, our source code and trained models are available for reproducibility and further studies.

\bibliographystyle{class/IEEEtran}
\bibliography{class/reference}
   
\end{document}